\documentstyle[prl,aps,epsf,twocolumn]{revtex}

\draft
\begin{document}

\title{Simultaneous Magneto-Optical Trapping of Two Lithium Isotopes
}
\author{Marc-Oliver Mewes, Gabriele Ferrari, Florian Schreck, Alice Sinatra, and Christophe Salomon}
\address{Laboratoire Kastler Brossel, Ecole Normale Sup\'erieure, 24 rue Lhomond, 75231 Paris CEDEX 05, France}
\date{\today}
\maketitle

\begin{abstract}
We confine $4 \times 10^8$ fermionic $^6$Li atoms simultaneously with
$9 \times 10^{9}$ bosonic $^7$Li atoms in a magneto-optical trap based
on an all-semiconductor laser system. We optimize the two-isotope
sample for sympathetic evaporative cooling. This is an essential step
towards the production of a quantum-degenerate gas of {\it fermionic}
lithium atoms.
\end{abstract}



The observation of Bose-Einstein condensation in atomic vapors
\cite{BEC} has made dilute {\it bosonic} quantum gases experimentally
accessible and the study of these systems has since been very
fruitful. Quantum degenerate gases of neutral {\it fermionic} atoms
such as $^6$Li or $^{40}$K have so far not been studied
experimentally. At temperatures below the Fermi temperature in such
systems energy and momentum transfer is modified by Fermi statistics
\cite{InelasticSuppression,FerrariProbeParticle}. One striking example
would be the partial suppression of spontaneous emission in an atomic
Fermi gas \cite{SpontaneousSuppression}. Also, a Fermi gas of neutral
atoms in a mixture of different hyperfine (HF) states might undergo a
BCS pairing transition and exhibit long range coherence and superfluid
behavior \cite{DiluteGasBCS}. It could give access to phenomena so far
only observed in strongly interacting Fermi systems such as atomic
nuclei, the quantum liquid $^3$He and degenerate electron gases.


Evaporative cooling of polarized atoms is so far essential for the
production of quantum degenerate gases \cite{KetterleDrutenReview}. It
is driven by elastic collisions. At ultra-low temperature $T$,
i.e.\,below a few millikelvin for lithium, collisions between bosons
or distinguishable particles are predominantly s-wave collisions while
Pauli exclusion prohibits s partial-waves of polarized fermions. In an
ultra-cold gas of indistinguishable fermions the elastic collision
rate diminishes proportional to $T^{-2}$ which has recently been
confirmed by De\ Marco {\it et al.} for $^{40}$K
\cite{JinSympathetic}. Fermionic atoms can be cooled {\it
sympathetically} by collisions in a mixture of different internal
states \cite{SympatheticCooling,JinSympathetic} or of different
species, yet to be implemented.

We intend to sympathetically cool fermionic $^6$Li with the bosonic
$^7$Li isotope. This could produce not only a quantum degenerate Fermi
gas but also a Bose-Einstein condensate of $^7$Li in both HF ground
states as well as a mixture of quantum degenerate gases of fermions
and bosons \cite{BoseFermiMixture}. One can also employ ultra-cold
bosons to probe collisional properties of a degenerate Fermi gas
\cite{FerrariProbeParticle}.

Previously several groups have studied samples of two atomic species
in a magneto-optical trap (MOT) \cite{TwoSpeciesStudies}. In this
article we describe the first realization of a MOT containing both
fermionic and bosonic lithium and its optimization for sympathetic
cooling. $4\times 10^8$ $^6$Li atoms and $9
\times 10^9$ $^7$Li atoms are simultaneously confined. These numbers together with
the density and temperature achieved should be sufficient to produce a
Fermi gas with a Fermi temperature on the order of $10\,\mu$K in a
harmonic magnetic trap of frequency $\overline{\omega}/2 \pi=
400$\,s$^{-1}$. The phase space density $\mathcal{D}$$=n_0\Lambda^3/f$
of unpolarized atoms ($f$ internal states) with a peak density $n_0$
and temperature $T$ is the number of identical atoms per cubic thermal
DeBroglie wave length $\Lambda=(2\pi\hbar^2/ m k_B T)^{1/2}$. The
achieved phase space density in the two-isotope magneto-optical trap
is $4\times10^{-6}$ for $^7$Li$(f=3)$ and $0.8\times10^{-6}$ for
$^6$Li$(f=2)$. In single isotope traps, the number of fermions(
$1.5\times 10^9$) exceeds the best previous realization of
laser-cooled fermions by more than one order of magnitude
\cite{JinSympathetic}. The number of trapped ${^7}$Li ($1.8
\times 10^{10}$) is also a factor of 10 improvement \cite{BestLithiumNumber}.

In future experiments, this two isotope sample will be polarized and
transferred into a magnetic trap. Bosonic lithium will be
evaporatively cooled. Fermionic lithium thermalizes with the bosons
via elastic collisions. Neglecting inelastic losses, all initially
confined fermions should reach the quantum degenerate regime. In
efficient evaporative cooling the phase space density $\mathcal{D}$ is
increased by $\sim 10^6$ by decreasing the atom number by $\sim 100$
\cite{KetterleDrutenReview}. For sympathetic cooling this implies that
$N$ initially confined $^7$Li atoms can sympathetically cool a sample
of $N/100$ $^6$Li atoms into the quantum degenerate regime
($\mathcal{D}$$\,\gg 1$). We therefore aim to maximize the number $N$
of $^7$Li atoms in the two-isotope trap while simultaneously confining
at least $N/100$ $^6$Li atoms.
 It is equally crucial that atoms thermalize
quickly during the trap lifetime. Thus the initial elastic collision
rate $\Gamma_i$ between $^7$Li atoms in the {\it mode-matched} and
compressed magnetic trap must be maximized. For a linear trapping
potential in three dimensions $\Gamma_{i}$ can be related to
quantities of the MOT as follows:
\begin{equation}
\label{GammaScaling}
\Gamma_{i} \propto N^{4/9} \mathcal{D}$$_i$$^{5/9} \propto N T^{-5/6}
\sigma^{-5/3}
\end{equation}
$N$ is the number of trapped atoms, $T$ is the temperature of the
sample and $\sigma$ the width of the gaussian density distribution
$n(r)=N/(\sqrt{2\pi}\sigma)^3 \times exp(-r^2/2\sigma^2)$ in the MOT.
We have optimized the laser-cooled sample of $^7$Li with respect to
$\Gamma_i$ in presence of $^6$Li.


In the experiment, the MOT is loaded from a Zeeman slowed lithium
beam. $^6$Li in the beam is enriched and has an abundance of about
$20\,\%$. Both isotopes are slowed and confined in the MOT with
$671$\,nm light near resonant with the D2 line, the
$2S_{1/2}\!\rightarrow\!2P_{3/2}$ optical transition. The isotopic
shift for this transition is $10$\,GHz. Each isotope requires two
frequencies to excite from the two HF ground states. The HF splitting
is $803.5$\,MHz for $^7$Li and $228.2$\,MHz for $^6$Li. Hence
simultaneous laser cooling of both lithium isotopes requires {\it
eight} different laser frequencies: four frequencies for Zeeman
slowing and four frequencies for magneto-optical trapping as shown in
Fig.\,\ref{fig:Frequencies}.

All frequencies are derived with acousto-optical modulators from two
grating-stabilized external-cavity diode lasers based on $30$\,mW
laser diodes. The lasers are frequency locked in saturated absorption
to the D2-lines of $^6$Li and $^7$Li respectively. The slowing light
is produced by geometrical superposition of the output of four
injection seeded $30$\,mW laser diodes. Four trapping frequency
components are geometrically superposed and $15$\,mW of this light is
injected into a tapered semiconductor amplifier chip. After spatial
filtering the chip produces up to $140$\,mW of trapping light
containing the four frequency components in a gaussian mode at
identical polarization as described in \cite{FerrariMOPA}. The
intensity ratio of the frequency components in the trapping beams can
be adjusted. This light is split up into six independent gaussian
trapping beams, each with maximum peak intensity of
$I_{max}=6$\,mW/cm$^2$, a $1/e^2$ intensity width of $3$\,cm and an
apertured diameter of $2$\,cm. The MOT is operated in a $4$\,cm$\times
4$\,cm$\times 10$\,cm Vycor glass cell of optical quality $\lambda/2$.
Background gas collisions limit the $1/e$ MOT lifetime $\tau$ to about
$25$\,s.

The trapped atom clouds of both isotopes are separately observed in
absorption imaging. For observation the trapping light and magnetic
field are switched off abruptly. The induction limited $1/e$ decay
time of the magnetic field is less than $50\,\mu$s. After free
ballistic expansion with an adjustable time-of-flight between
$150\,\mu$s and $7\,$ms the sample is illuminated for $80\,\mu$s by a
probe beam. This probe excites either $^7$Li from the F=2 ground state
or $^6$Li from the F=3/2 ground state to the $2P_{3/2}$ excited state
manifold. The absorption shadow of the sample is imaged onto a
charge-coupled device (CCD) camera. A separate repumping beam which is
not projected onto the camera excites atoms in the other HF ground
state to avoid optical pumping. The density distribution, atom number
and temperature of the sample are obtained from absorption images for
different ballistic expansion times.


Both isotopes are magneto-optically trapped in two steps: In the first
step, {\it the loading phase}, the capture volume and velocity of the
trap is large such that the number of trapped atoms is maximized. In
the second step, {\it the compression phase}, the already trapped
atoms are compressed in phase space such that the initial elastic
collision rate $\Gamma_i$ is maximized.

All four frequencies $\nu_{P7},\,\nu_{R7},\,\nu_{P6}$ and $\nu_{R6}$
of the MOT are exciting on the D2-line. We maximized the number of
$^7$Li atoms and $^6$Li atoms in separate MOTs as well as the number
of $^7$Li atoms in the two-isotope trap. The maximization involved the
detunings $\delta_{P7},
\delta_{R7}, \delta_{P6}, \delta_{R6}$ of the light components from the
cooling and repumping transitions of the two isotopes
(Fig.\ref{fig:Frequencies}), the intensities of all frequency
components and the strength of the magnetic field of the MOT.

First, the atom number in separate {\it single-isotope} MOTs with only
the two frequencies for the respective isotope present was optimized.
We were able to capture up to $1.8\times10^{10}$ $^7$Li atoms and
$1.5\times10^{9}$ $^6$Li atoms. The atom number $N$, peak density
$n_0$, temperature $T$ and the respective detunings are listed in
Tab.\ref{tab:MOTtable}. The atom number is accurate to within a factor
of 2 and this dominates the uncertainty in the density determination.
The temperature uncertainty is $0.2$\,mK. For both isotopes the atom
number is maximized at large frequency detunings and equal intensities
in both frequency components. The optimum magnetic field gradient $B'$
along the symmetry axis of the magnetic quadrupole field of the MOT is
about $35$\,G/cm for both isotopes. The MOT was operated at maximum
intensity $I_{max}=6$\,mW/cm$^2$ in each of the six beams.

In the $^7$Li trap, at low atom number ($\leq 10^9$) the temperature
is the Doppler temperature (1.1 mK at $\delta_{P7}=-8\Gamma$), as
shown in Fig\,\ref{fig:Ndependence}. At large atom numbers, for
$5\times 10^9$ to $2\times 10^{10}$ trapped atoms, the temperature is
$1.5(2)$\,mK and nearly constant while the density typically saturates
at $n_0$=$3 \times 10^{11}$\,cm$^{-3}$. In this regime the number of
trapped $^7$Li atoms is limited by loss due to inelastic radiative
escape (RE) or fine structure changing (FS) collisions. In
steady-state, the slow atom flux $\mathcal{F}$$=2\times10^9$\,s$^{-1}$
is balanced by the trap loss according to
\begin{equation}
\label{NLimit}
\mathcal{F}$$ = N/\tau + \beta n_0 N/\sqrt{8}
\end{equation}
$\tau=25$\,s is the background gas limited lifetime of the MOT. The
two-body loss coefficient $\beta=6\times 10^{-13}$\,cm$^3/$s was
experimentally determined and is consistent with previous studies of
trap loss in a $^{7}$Li MOT \cite{LiMOT}.

In $^7$Li the HF splitting between the F$'$=3 and the F$'$=2 excited
states is $1.6\,\Gamma$ and in $^6$Li, $0.5\,\Gamma$ between the
F$'$=5/2 and the F$'$=3/2 excited states, where $\Gamma=5.9\,$MHz is
the natural width of the lithium D-lines. Despite the inverted excited
state HF structure of both lithium isotopes, the small HF splitting
leads to off-resonant excitation of the F=2$\rightarrow$F$'$=2
transition in $^7$Li, and the F=3/2$\rightarrow$F$'$=3/2 transition in
$^6$Li and frequent decay into the lower HF ground state. The
repumping light component is therefore of equal importance as the
principal trapping light. In fact, we only obtained a MOT with the
repumping light also in a six-beam MOT configuration. This is not
required in MOTs of other alkalis with larger HF splitting, such as
Cs, Na or Rb.

For the two-isotope trap, the $^6$Li repumping transition
F=1/2$\rightarrow$F$'$=3/2 is about $7(3)\,\Gamma$ to the blue of the
F=2$\rightarrow$F$'$=1 resonance in the D1 line of $^7$Li. If both
lithium isotopes are simultaneously confined, the $^6$Li repumping
light component $\nu_{R6}$ frequently excites this {\it non-cooling}
transition and significantly weakens the confinement of the trap for
$^7$Li. This leads to smaller number of trapped $^7$Li atoms in the
presence of $^6$Li light. We reduce this harmful effect by detuning
towards the D1 resonance while reducing the intensity. The coincidence
could also be avoided by repumping $^6$Li on the D1 line instead.
Aside from the light induced trap loss we do not observe mutual
effects due to the presence of both isotopes (such as collision
induced trap loss, heating or a modification of the spatial
distribution).

As discussed in the introduction, for efficient sympathetic cooling in
subsequent stages of the experiment we require about $100$ times less
precooled $^6$Li atoms than $^7$Li atoms. We decrease the intensity of
$^6$Li light in the two-isotope trap in order to reduce the $^7$Li
trap loss. This maximizes the number of $^7$Li atoms at the cost of
less $^6$Li atoms. We are able to confine $9\times 10^{9}$ $^7$Li
atoms together with $4\times 10^{8}$ $^6$Li atoms. This result was
obtained with an intensity relation between the four frequency
components $\nu_{P1},\,\nu_{R1},\,\nu_{P2}$ and $\nu_{R2}$ of
$8\,:\,8\,:\,2\,:\,1$ and a magnetic field gradient $B'=35$\,G/cm.
$N$, $n_0$, $T$ and the respective detunings for the two-isotope MOT
are listed in Tab.\,\ref{tab:MOTtable}.


After loading the trap it is possible to further compress the sample
in phase space and maximize the initial elastic collision rate
$\Gamma_{i}$ by changing the laser parameters for the duration of a
few milliseconds. From Eq.\ref{GammaScaling} follows that in case of
{\it no} loss of atoms during compression a maximization of
$\Gamma_{i}$ also maximizes $\mathcal{D}$$_{i}$. For the compression,
we optimize $\Gamma_{i}$ with respect to the total laser intensity and
the frequency detunings $\delta_{P6},\,\delta_{P7},\,
\delta_{R6},\,\delta_{R7}$ while keeping $B'$ constant at
$35$\,G/cm. We compress the single-isotope MOTs as well as the
two-isotope sample. For sympathetic cooling we are especially
interested in maximizing $\Gamma_i$ for $^7$Li in the presence of
$^6$Li. As shown in Tab.\,\ref{tab:CMOTtable}, decreasing
$\delta_{R7}$, i.\,e. detuning $\nu_{P7}$ further to the red of the
transition, and increasing $\delta_{P7}$ towards resonance while
reducing the overall laser intensity to
$0.3\,I_{max}$=$1.8$\,mW/cm$^2$ results in a $40 \%$ drop in
temperature and increases the density by $70\,\%$. $30\,\%$ of the
initially confined atoms are lost during the first 3\,ms of
compression. This loss is either due to FS- and RE-collision induced
heating during the initial compression stage or due to a loss of parts
of the cloud as a consequence of an abrupt change of laser frequencies
(see Fig.\,{\ref{fig:temporaldynamics}). According to
Eq.\ref{GammaScaling} compression increases $\Gamma_i$ by $60\%$.
After compressing the two-isotope trap for $3$ ms, we obtain a maximum
of $6\times 10^{9}$ $^7$Li atoms at a peak density of $4\times
10^{11}$\,cm$^{-3}$ and a temperature of $0.6$\,mK together with
$^6$Li at a density $6.5\times 10^{10}$ cm$^{-3}$ and a temperature of
$0.7\,$mK. Aside from the initial loss during the first few
milliseconds, the $1/e$ trap lifetime also decreases to about $30$\,ms
for both isotopes.

Fig.\ref{fig:temporaldynamics} shows the typical temporal dynamics of
atom number, width and temperature of the ${^7}$Li compressed MOT
(CMOT) for the first $6.5$\,ms of compression. $\Gamma_i$ and
$\mathcal{D}$$_i$ reach a maximum after about $3$\,ms. The CMOT is
rather insensitive to the repumping frequency $\nu_{R7}$: detuning by
$2\,\Gamma$ above and below the optimized value of $5.8\,\Gamma$
decreases $\Gamma_i$ by less than $25\%$. The CMOT is much more
sensitive to the detuning $\nu_{P7}$ of the trapping light. We
optimized $\Gamma_i$ with respect to the duration $t$ of the
compression phase and the detuning. The maxima of $\Gamma_i$ and
$\mathcal{D}$$_i$ remain at a constant value for $t\ge 3$\,ms but
shift to different detuning parameters with increasing CMOT duration.


To summarize, we showed that it is possible to trap about $6\times
10^{9}$ $^7$Li atoms together with $3\times 10^{8}$ $^6$Li atoms in
the two-isotope MOT at at phase space densities of $\sim 10^{-6}$.
These results are comparable to results achieved with single isotopes
of Na or Cs in a {\it dark} SPOT \cite{darkSPOT}. In combination with
a strong confining magnetic trap we expect an initial elastic
collision rate well above $10\,$s$^{-1}$, despite the small triplet
scattering length of $1.4\,$nm ($^7$Li-$^7$Li) and $2.0$\,nm
($^6$Li-$^7$Li)\cite{HuletScatteringlength}. This can lead to the
production of quantum degenerate Bose and Fermi gases of lithium by
forced evaporation and sympathetic cooling within a few seconds.

We are grateful for experimental assistance of Fabrice Gerbier. We
thank C. Cohen-Tannoudji and J. Dalibard for discussions and Kirk
Madison for careful reading of the manuscript. M.-O.\,M. and F.\,S.
are supported by a Marie-Curie Research fellowship of the EC and a
doctoral fellowship from the DAAD(HSP 3) respectively. This work was
partially supported by CNRS, Coll\`ege de France, DRED, and the EC
(TMR Network No. ERB FMRX-CT96-0002). Laboratoire Kastler Brossel is
{\it Unit\'e de recherche de l'Ecole Normale Sup\'erieure et de
l'Universit\'e Pierre et Marie Curie, associ\'ee au CNRS}.

\begin{table}

 \begin{tabular}{l|c c|c c}
                                &\multicolumn{2}{c}{single isotope MOT}&\multicolumn{2}{c}{two isotope MOT}\\[0.5ex]

                                &$^7$Li                &$^6$Li                 &$^7$Li                 &$^6$Li\\ [0.5ex]\hline

 N                              &$1.8\times 10^{10}$   &$1.5\times 10^{9}$     &$9\times 10^{9}$     &$4\times 10^{8}$\\
 n\,$\-[$cm$^{-3}\-]$           &$3\times 10^{11}$   &$1.0\times 10^{11}$    &$2.5\times 10^{11}$    &$5\times 10^{10}$\\
 T\,$\-[$mK$\-]$                &$1.5$                 &$0.7$                  &$1.0$                  &$0.7$\\
 $\delta_{P7,6}\,\-[\Gamma\-]$  &$-8.0$                &$-2.7$                 &$-8.0$                 &$-2.7$\\
 $\delta_{R7,6}\,\-[\Gamma\-]$  &$-5.8$                &$-5.1$                 &$-5.8$                 &$-5.1$\\
 \end{tabular}

 \caption{
 \label{tab:MOTtable}
  Comparison of atom number $N$, peak density $n_0$, temperature $T$
  and frequency detunings for the single-isotope and two-isotope
  MOT.}

\end{table}

\begin{table}
 \begin{tabular}{l|c c|c c}

                                &\multicolumn{2}{c}{single isotope CMOT}&\multicolumn{2}{c}{two isotope CMOT}\\[0.5ex]

                                &$^7$Li                &$^6$Li                  &$^7$Li                 &$^6$Li\\ [0.5ex]\hline

 N                              &$7\times 10^{9}$    &$5\times 10^{8}$       &$6\times 10^{9}$     &$3\times 10^{8}$\\
 n\,$\-[$cm$^{-3}\-]$           &$4\times 10^{11}$   &$1.5\times 10^{11}$    &$4\times 10^{11}$  &$6.5\times 10^{10}$\\
 T\,$\-[$mK$\-]$                &$0.6$             &$0.4$              &$0.6$            &$0.7$\\
 $\delta_{P7,6}\,\-[\Gamma\-]$  &$-3.0$                &$-2.7$                 &$-3.0$               &$-2.7$\\
 $\delta_{R7,6}\,\-[\Gamma\-]$  &$-9.0$                &$-2.7$                 &$-9.0$               &$-5.8$\\
 \end{tabular}

 \caption{
 \label{tab:CMOTtable}
  Comparison of atom number $N$, peak density $n_0$, temperature $T$
  and frequency detunings for the single-isotope and two-isotope compressed MOT.}

\end{table}

\begin{figure}[h]

\caption{
\label{fig:Frequencies}
Frequencies employed to slow (dashed arrows) and magneto-optically
trap (solid arrows) both lithium isotopes. The detunings of the
frequencies from the respective resonances are marked with a dotted
line. The detuning of the slowing light from the respective zero
magnetic field transitions for $^7$Li is $-426$\,MHz and $-447$\,MHz
for $^6$Li.}
\end{figure}

\begin{figure}[h]

\caption{
\label{fig:Ndependence}
Temperature $T$, 1D {\it rms} width $\sigma$ and peak density $n_0$ in
a $^7$Li MOT versus atom number $N$. $N$ was varied by changing the
loading time. In b) the width along the symmetry axis z of the
magnetic field (solid data points) is $\sim 40\%$ smaller than in the
radial direction (hollow points) as expected from a simple MOT
model($1/\sqrt{2})$.}
\end{figure}

\begin{figure}[h]

\caption{
\label{fig:temporaldynamics}
Temporal dynamics of the compression phase of a $^7$Li MOT. Atom
number $N$, temperature $T$, and {\it rms} width $\sigma_z$ after
abrupt change of the laser parameters at $t=0$.}
\end{figure}

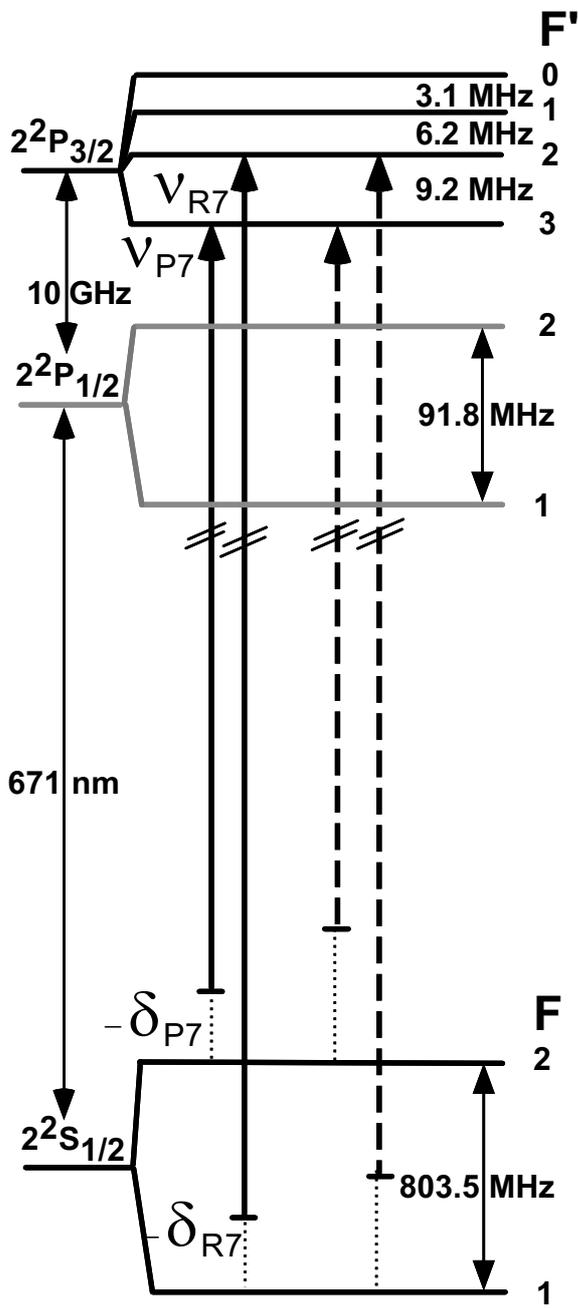
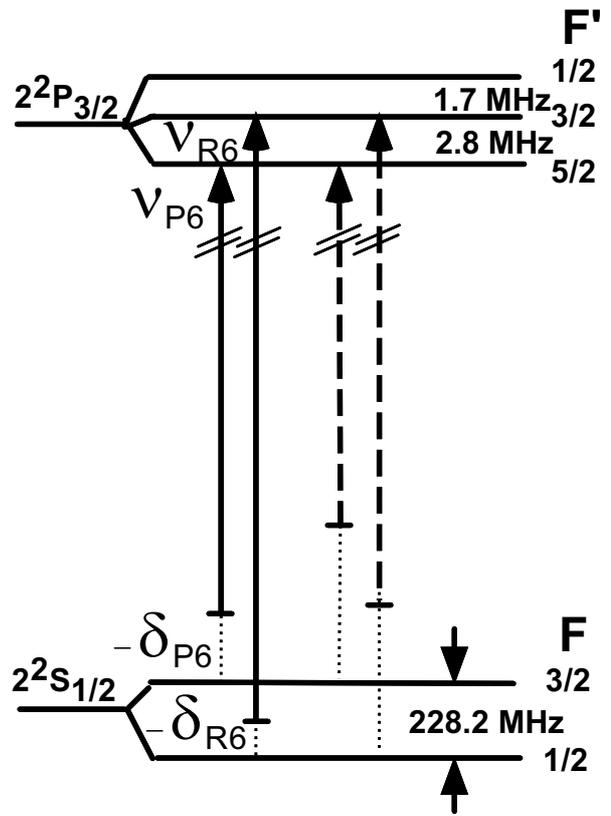
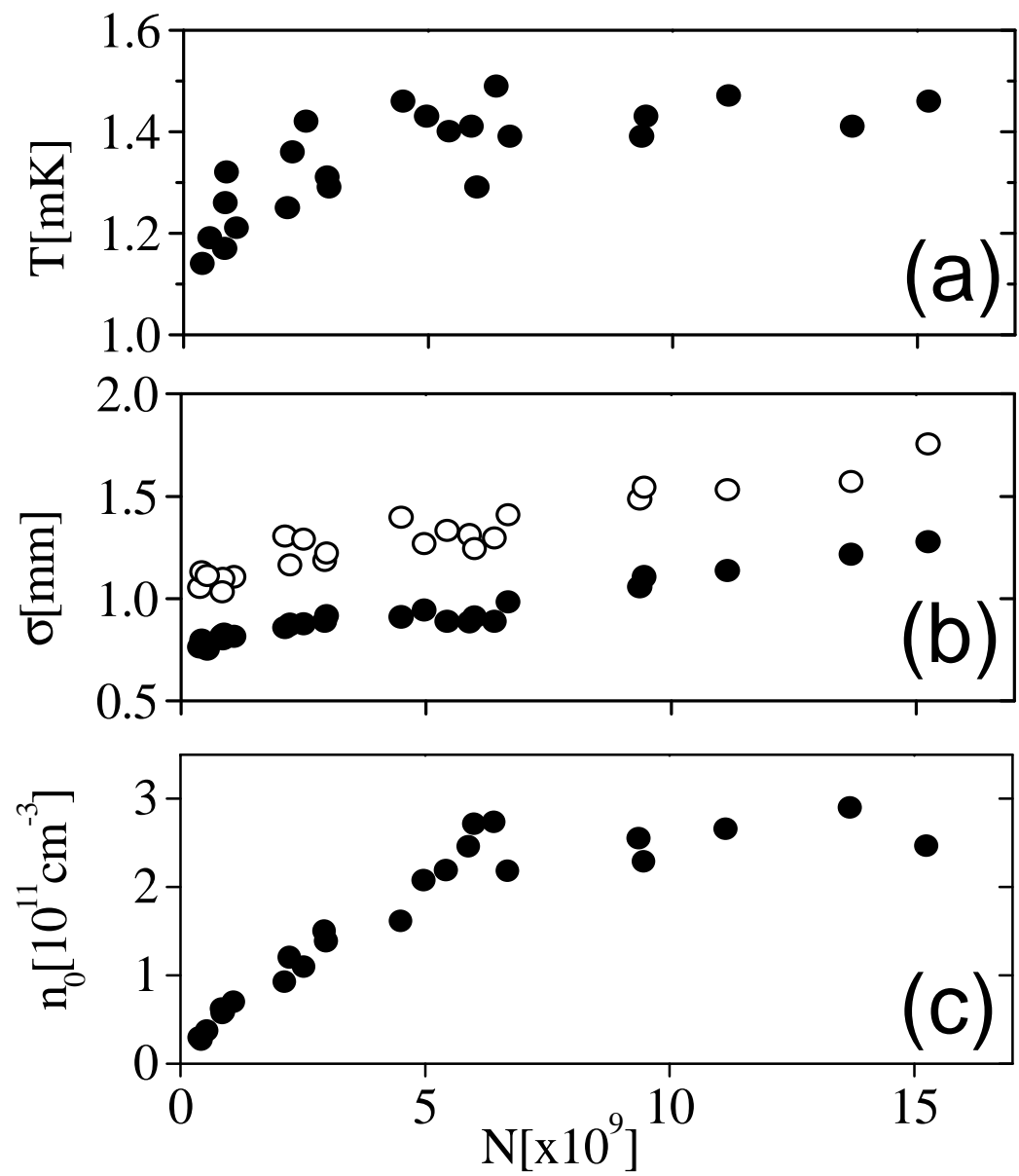
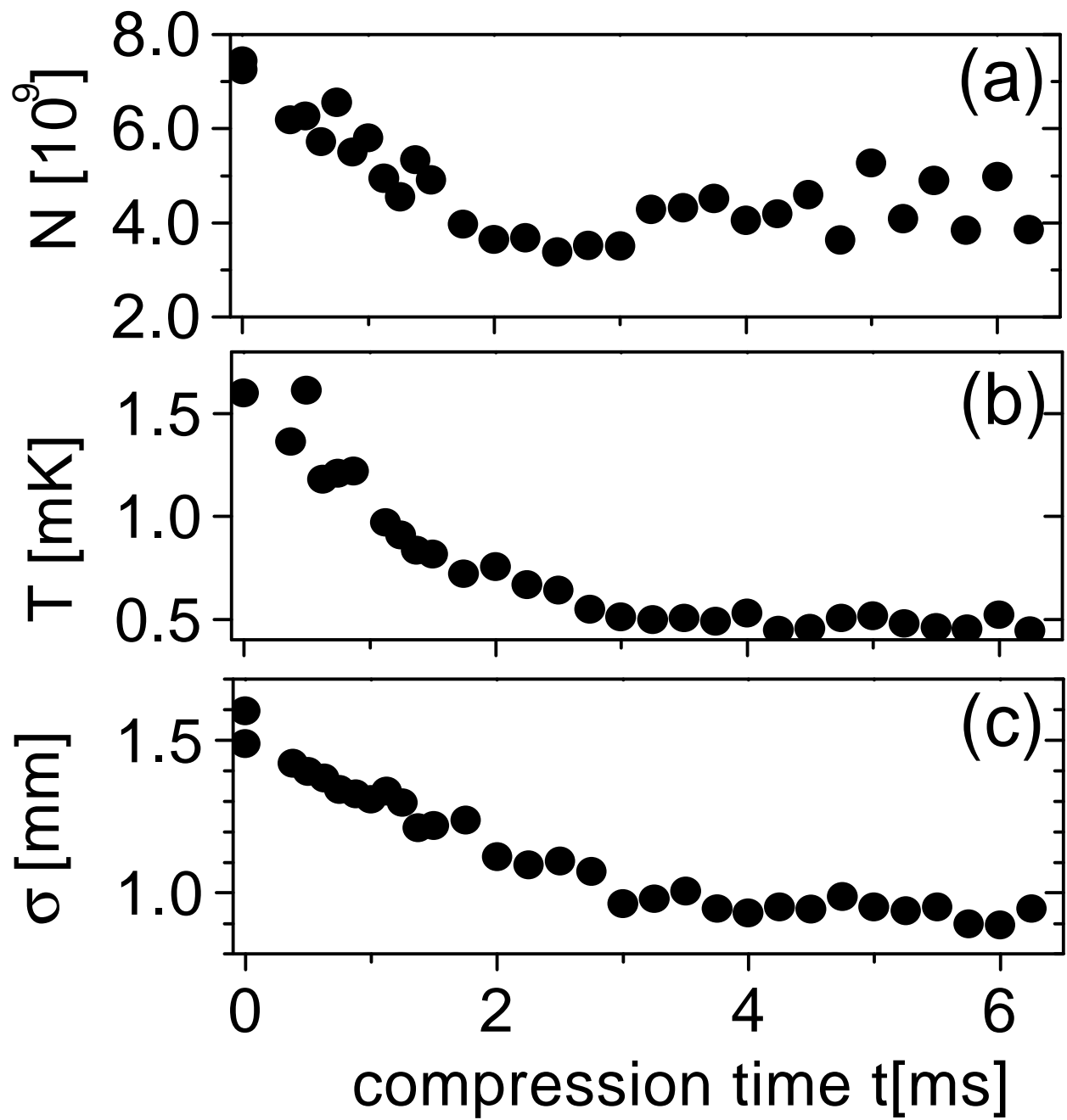
\end{document}